# Vacuum for Accelerators: Introduction to Materials and Properties

*S. Sgobba*
CERN, Geneva, Switzerland

**Abstract**
In modern accelerators, stringent requirements are placed on the materials used for vacuum systems. Their physical and mechanical properties, machinability, weldability and brazeability are key parameters. Adequate strength, ductility, magnetic properties at room as well as low temperatures are important factors for vacuum systems of accelerators working at cryogenic temperatures. In addition, components undergoing baking or activation of Non-Evaporable Getters (NEG) or directly exposed to the beam impose specific choices of material grades for suitable outgassing and mechanical properties in a large temperature range. Today, stainless steels are the dominant materials of vacuum systems. The reasons for specific requirements in terms of metallurgical processes are detailed for obtaining adequate purity, inclusion cleanliness and fineness of the microstructure. In many cases these requirements are crucial to guarantee the final leak tightness of the vacuum components. Innovative manufacturing and material examination technologies are also treated.

## 1     An historical introduction

In 1936, Springer in Berlin published a treaty handbook on 'Werkstoffkunde der Hochvakuumtechnik' by Espe and Knoll [1], which saw a fivefold increase in contents by the time of its first English edition in 1966 [2]. The Espe handbook is the first systematic and comprehensive presentation of high vacuum materials. Espe pointed out that the selection and handling of vacuum materials is based on a different viewpoint than that for ordinary industrial construction, mentioning the ease of degassing, an adequate strength at high as well as low temperature, matching values of thermal expansion coefficients, and purity of materials as principal factors to be considered in material selection for vacuum applications. Ease of fabrication and cost of vacuum materials were considered as being often "of secondary importance". Very constant properties of the raw materials, which should be specially prepared, exact knowledge of the material properties, critical selection, and careful controls were mentioned as aspects of paramount importance. Among the materials treated, glass has a primary importance, followed by ceramics, precious metals (mainly Pt used in glass-metal sealing), iron and steels (little attention is devoted in the handbook to stainless steels), refractory metals and alloys, polymers, Al, Cu, Ni and their alloys. At that time, electrode systems and vacuum tubes were the main applications and formed the first background to this voluminous treaty of several thousand pages.

A very comprehensive handbook on materials is the 'Handbook of Materials and Techniques for Vacuum Devices' by Kohl [3], a revision and expansion of its predecessor 'Materials and Techniques for Electron Tubes' [4]. Glass and ceramics are the materials of main interest, together with metallic materials. Additional handbooks containing more or less comprehensive reviews of materials for high vacuum were published in the 1960s [5−7]. In a recent book by O'Hanlon [8], containing a chapter on materials, stainless steels receive the largest attention compared to other metallic materials. The very recent second edition of the Jousten handbook [9] provides similar recommendations, mentioning also corrosion resistance as an important aspect of material selection. A comprehensive review of materials for vacuum technology, of which this article is an update, is provided in Ref. [10]. Today austenitic stainless steels represent the reference material for many vacuum devices, due to their corrosion resistance, strength and ductility retained at ambient and service temperatures, suitability for intended cleaning procedures, stability of properties in service, toughness, magnetic properties, sharpness (retention of cutting edges in applications to vacuum seals), rigidity, and dimensional stability [11]. Aspects linked to the steelmaking and processing are treated, including possible sensitization issues linked to vacuum firing or stress relieving.

In addition to stainless steels, other families of materials relevant for vacuum applications in accelerators that are covered in the present overview are Al and alloys, Cu and alloys, including melt spun and oxide dispersion strengthened alloys, respectively. Special factors will be discussed that are relevant to manufacturing techniques for vacuum purposes such as welding, including laser and Electron Beam (EB) welding, diffusion and explosion bonding. Finally, innovative technologies, processes and inspection techniques will be presented such as near net shaping techniques by Hot Isostatic Pressing (HIP) and X-ray microtomography, allowing the metrological and metallurgical inspection of complex shape components, materials, and welds. For each material family and process, examples of applications are given. Failure analyses, including corrosion issues are discussed.

## 2 Stainless steels

Stainless steel plays a crucial role in the construction of modern accelerators. The example of the dipole and quadrupole magnets of the Large Hadron Collider (LHC) is representative in this respect. A special austenitic stainless steel was developed [12] for the beam screen and the cooling capillaries of the machine vacuum system, retaining high strength, ductility, and low magnetic susceptibility at the working temperature between 10 K and 20 K. Several tens of kilometres of components have been produced in this special grade that will be used again for the beam screen of the High Luminosity LHC (HL-LHC) project (3.1 km of finished strip for the beam screen and 4.6 km of seamless cold-drawn cooling tubes in lengths of up to 14 m). The LHC magnet cold bore is manufactured as a seamless 316LN tube. 316LN is also the grade of the shrinking cylinder of the dipole magnets. For the LHC, 2500 bent plates of 15.35 m length and 10 mm thick for a total weight of approx. 3000 t were used, welded longitudinally by a special Surface Tension Transfer (STT) technique combined with traditional pulsed Metal Inert Gas (MIG) welding. More than 2800 magnet end covers of complex shape, including several nozzles, have been fabricated for the LHC, starting from HIPed 316LN powders and near net shaped into geometry close to the final form. Between the magnets, some 1600 interconnections consist of several thousand of leak tight components, mainly working at cryogenic temperature (1.9 K). Interconnection components are also essentially based on austenitic stainless steels. For the convolutions of the several thousands of bellows involved in the machine and working under cyclic load at 1.9 K, a special remelted 316L grade was used showing an extremely low inclusion content and improved austenite stability at the working temperature. The grade is highly formable at room temperature (RT) [13].

### 2.1 Families of stainless steels

Stainless steel can be defined as a ferrous alloy containing a minimum of 12 % Cr [14], with or without other elements [11,15]. Chromium, as well as additional alloying elements, imparts corrosion and oxidation resistance to steel. Figure 1 is an iron-chromium diagram, which is the foundation of stainless steels [14]. On the 100% Fe axis of the diagram, one recognizes the stability domains of the various phases of iron as a function of temperature: $\alpha$- and $\delta$-iron, corresponding to the ferromagnetic ferritic phase of body centred cubic (bcc) structure, which are present up to 912°C and in the ranges between 1394°C and 1538°C, respectively, and $\gamma$-iron, corresponding to the austenitic phase of face centred cubic structure (fcc), in the range between 912°C and 1394°C. This phase is non-ferromagnetic. β-iron is an obsolete designation of paramagnetic α-iron above the Curie temperature which is not really a distinct phase.

In the diagram of Fig. 1 [16], the ferritic phase is extensive while the γ phase is limited to a loop. This diagram is the basis for identifying the two first families of stainless steels: ferritic types, with Cr contents between 14.5% and 27%, and martensitic types, that are iron-chromium steels with small additions of C and other alloying elements, usually containing no more than 14% Cr (excepting some types with 16% to 18% Cr) and sufficient C to promote hardening. From the diagram it can be seen that ferritic grades do not transform to any other phase up to the melting point, and hence can only be strengthened by cold working. Martensitic grades, due to the addition of C to the Fe-Cr system enlarging the domain of stability of the γ-phase, can be hardened by a rapid cooling in air or a liquid medium from above a critical temperature. This results in grades that have excellent strength.

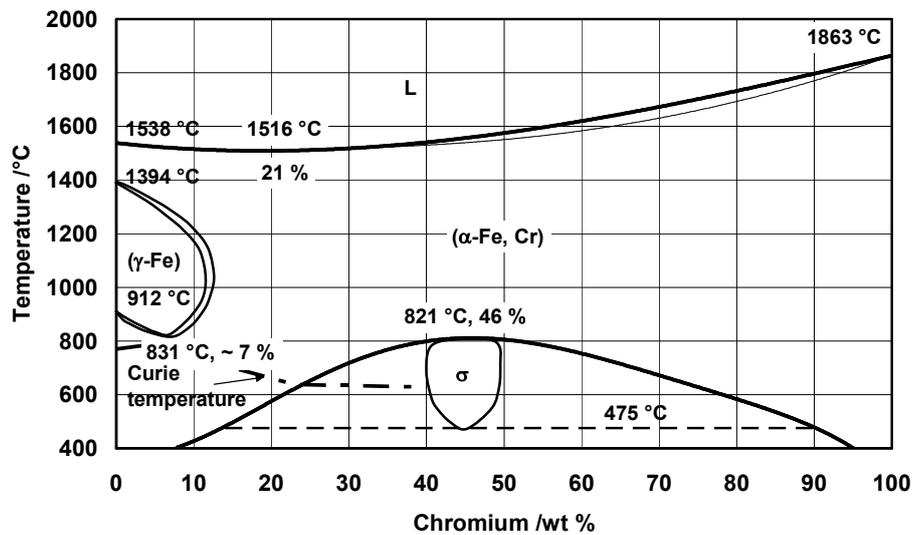

**Fig. 1:** The Fe-Cr phase diagram (from [16])

Ferritic and martensitic grades are generally inferior, in terms of corrosion resistance, to austenitic grades. They are ferromagnetic and undergo ductile-to brittle transition at cryogenic temperatures. Moreover, at temperatures in the range of 1000°C or more (solution annealing), ferritic grades undergo grain growth, while martensitic grades are subject to a loss of the martensitic hardening phase due to a reversion of martensite to austenite in the annealing conditions associated with a high temperature treatment. On the other hand, austenite is non-magnetic, does not undergo any ductile-to-brittle transition below RT, and is less subject to grain growth during vacuum firing. For these reasons, the austenitic grades are a first choice for vacuum applications. Austenitic grades, due to their high Cr and Ni content, are the most resistant to corrosion of the stainless steel family [14]. Because of their limited application in vacuum devices, precipitation-hardening grades are not mentioned here.

## 2.2 Austenitic stainless steels

These are formed by the addition of elements (Ni, Mn, N…), to the Fe-Cr system of Fig. 1, broadening the domain and enhancing the stability of the γ phase. For sufficient additions of a balanced amount of adequate alloying elements, the formation of ferrite can be suppressed and the tendency to form martensite on cooling or during work hardening can be partially or completely suppressed. As an example, the addition of 8% to 10% Ni to a low C FeCr steel can already allow a relatively stable austenitic structure at RT to be obtained. The '304-type' stainless steel (18Cr8Ni, on an Fe basis) is a typical example of a very common iron-chromium-nickel austenitic stainless steel. This grade is generally applied in its low (L) carbon version 304L (C ≤ 0.030%), showing enhanced corrosion resistance and ductility especially in welded structures.

Figure 2 is a pseudobinary section of the FeCrNi ternary diagram for increasing Cr+Ni contents [17]. In this diagram, for a total Fe of 70%, it is straightforward to visualize the basic 304 type of austenitic stainless steel. Continuous lines in the diagram correspond to real transformations, dashed lines to transformations not observed in practical conditions. For a 20% Cr content and 10% of Ni (composition corresponding to the far left dashed vertical line), the diagram shows that cooling from the usual solution annealing temperature of 1050°C, can result in some residual ferrite. Indeed, 304L grade retains, when cooled from solution annealing (or precipitates during a welding operation), a given amount of untransformed δ-ferrite. Increasing Ni and reducing Cr (vertical lines to the right) enhances stability of γ-phase against precipitation of ferrite and, for sufficient Ni, can totally suppress it.

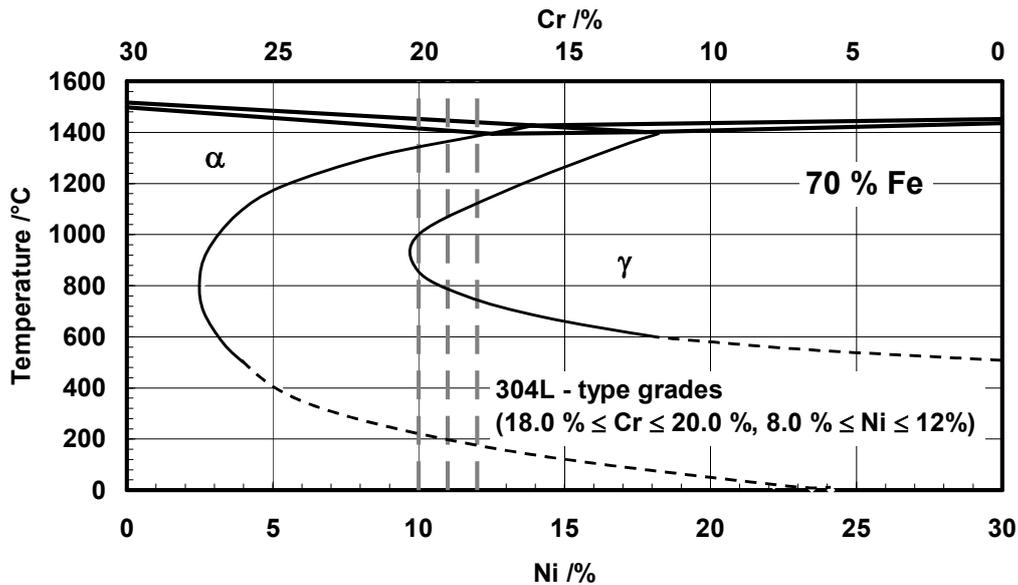

**Fig. 2:** Cross-section of the Fe-Cr-Ni ternary phase diagram [17]

From the diagram in Fig. 2, it appears clear that:

1) austenitic stainless steels may contain residual amounts of δ-ferrite, which can be critical due to its reduced toughness and to its ferromagnetic nature for specific vacuum applications, particularly for components that have to be applied at cryogenic temperature and/or in critical magnetic environments;

2) elements like Ni extend the domain of austenite, while Cr reduces it. Elements such as C, N, Mn (up to some extent)… or a combination of them play a role similar to Ni ('Ni-equivalents', $Ni_{eq}$), while elements like Mo, Si, Nb… or their combination act as 'Cr-equivalents' ($Cr_{eq}$). Schaeffler, DeLong, Hull, or Espy diagrams allow the total ferrite formation effects to be predicted in a multielement system, such an industrial stainless steel grade and its welds (Section 2.5.1).

## 2.3 Grades of practical interest for vacuum applications

The first four rows in Table 1 show the composition of the grades of stainless steels (304L, 316L, 316LN) of main interest for vacuum applications and covered by CERN specifications. The mentioned composition ranges are based on the last version of the CERN specifications [18−25], with reference to the relevant standard grade designations.

**Table 1:** Grades of practical interest for vacuum applications. Composition ranges according to CERN specifications, wherever applicable, or EN 10088 [26]. Single values are maximum admitted values.

| Grade (AISI, or 'commercial designation') | Grade (EN, symbolic and numeric) | C | Cr | Ni | Mo | Si | Mn | N | Others |
|---|---|---|---|---|---|---|---|---|---|
| 304L [18,19] | X2CrNi19-11 1.4306 | 0.030 | 17.00–20.00 | 10.00–12.50 | | 1.00 | 2.00 | | P≤0.030, S≤0.015 Co≤0.10 |
| 316L [20,21] | X2CrNiMo18-14-3 1.4435 | 0.030 | 17.00–19.00 | 12.50–15.00 | 2.50–3.00 | 1.00 | 2.00 | | P≤0.030, S≤0.015 Co≤0.10 |
| 316L for bellows [22] | X2CrNiMo18-14-3 1.4435 | 0.030 | 16.50–19.00 | 13.50–16.00 | 2.50–3.00 | 1.00 | 2.00 | 0.050 | P≤0.030, S≤0.010 Co≤0.10 |
| 316LN [23−25] | X2CrNiMoN17-13-3 1.4429 | 0.030 | 16.00–18.50 | 12.00–14.00 | 2.00–3.00 | 1.00 | 2.00 | 0.14−0.20 | P≤0.045, S≤0.015 Co≤0.10 |
| 316Ti [26] | X6CrNiMoTi17-12-2 1.4571 | 0.08 | 16.50–18.50 | 10.50–13.50 | 2.00–2.50 | 1.00 | 2.00 | | P≤0.045, S≤0.015, 5xC≤Ti≤0.70 |
| 'P506' [12] | - | 0.030 | 19.00–19.50 | 10.70–11.30 | 0.80–1.00 | 0.50 | 11.80–12.40 | 0.30–0.35 | P≤0.020, S≤0.002, B≤0.002, Cu≤0.15, Co≤0.10 |

AISI 304L is a general purpose grade. For vacuum applications, the grade should be purchased through a careful specification, aimed at achieving a substantially austenitic microstructure and a controlled maximum level of inclusions (Section 2.4). Even in its higher alloy version 1.4306 specified by CERN, due to the limited amount of alloying elements, its magnetic susceptibility can be subject to increase by martensitic transformation. This can occur upon cooling to RT or to cryogenic temperatures, and following work hardening at RT or lower [27]. Its selection should be carefully considered in applications where an increase of magnetic susceptibility might be of concern. The price of this grade is approximately 3.5 EUR/kg (at mid-2017 rates) for a general purpose version, and 6 EUR/kg for a vacuum specified wrought product.

AISI 316L is a Mo bearing grade. Mo enhances corrosion resistance and austenitic stability versus martensitic transformation. The ferrite-promoting characteristics of Mo (Section 2.2 and Section 2.5.1.1) have to be compensated by adjustments in Cr and Ni to achieve an almost fully austenitic microstructure. Due to its formability, ductility, and increased austenitic stability compared to 304L, this grade is covered by a special CERN specification applicable to the material of the bellows convolutions of the LHC interconnections. In the form of thin sheet for these convolutions, prices rise up to 50−80 EUR/kg. The high alloy EN 1.4435 version is preferred to EN 1.4404 for vacuum applications, featuring a composition more stable against martensitic transformations and in general also against δ-ferrite precipitation in welds. Wrought 316L products for vacuum applications, newly covered by CERN specifications [20,21], have prices in between 304L and 316LN for equivalent steelmaking and metalworking processes and inspections.

AISI 316LN is a nitrogen bearing stainless steel. N increases austenite stability against martensitic transformations and is a powerful austenite former with respect to ferrite. N substantially increases strength, while allowing ductility to be maintained down to cryogenic temperatures [28]. Due to limited softening compared to 304L and 316L, 316LN is the grade preferentially selected when vacuum firing is required. Prices in the range of 11 EUR/kg (bars) to 32 EUR/kg (plates) are common for wrought products issued from Electroslag Remelted (ESR) ingots. At the level of ingot remelting and for steelmaking companies equipped with last generation, highly productive ESR units, the additional cost of ESR applied to austenitic stainless steel grades can be very limited, of the order of 1 EUR/kg [29]. However, additional value is brought by redundant multidirectional forging of the ingots. Open die forged ESR 316LN products (Section 2.4) have prices of the order of 50 (but up to above 100) EUR/kg.

AISI 316Ti is an example of a Ti 'stabilized' grade. Stabilized grades contain higher C than low C grades (limited to 0.030% C max). They are alloyed with Nb, Ta or Ti to prevent carbide precipitation. AISI 316Ti, or similar stabilized grades (AISI 321, AISI 347), are offered by steel suppliers or component producers as an alternative to low carbon grades. Stabilized grades should be generally avoided for demanding vacuum applications, since the addition of the stabilizer elements results in precipitate carbides, reducing the cleanliness of the steel (Section 2.4) and toughness in specific low temperature applications.

P506 is a stainless steel specially developed by CERN [12], belonging to the family of high Mn, high N stainless steels and successfully produced by Böhler/AT and Aubert et Duval/FR (under a different designation). This special composition allows low relative magnetic permeability (<1.005) to be maintained down to cryogenic temperatures in the base material and in its welds. A full stability of the austenite versus δ-ferrite precipitation in welds, and/or versus martensitic transformations, even when deformed at very low temperatures, is guaranteed by the alloy composition [28]. Its price (50−80 EUR/kg) is comparable to that of 316L sheets or multidirectionally forged products in 316LN grade.

## 2.4 Special requirements on the microstructure of stainless steels for vacuum applications

Figure 3 shows an end fitting welded to a thin bellows convolution, which was intended for application in the vacuum system of the Compact Muon Solenoid (CMS) experiment. The bellow ends are machined from stainless steel forged bars, showing strings of non-metallic inclusions aligned parallel to the axis of the bar (the primary metal flow direction). These inclusions result in a leak through the material thickness of some $10^{-4}$ mbar·l·s$^{-1}$ [30].

Several leaks through the base material thickness of flanges of the plug in modules of LHC magnet interconnections have been reported and investigated [31]. In this case, leaks are due to exogenous macroinclusions (Fig. 4). Macroinclusions are understood in terms of the presence of entrapped slag or refractories added to protect the steel melt during the melting phases, segregated during the solidification and not fully removed due to an insufficient top discard of the steel ingot. Slag trapped in the ingot results in elongated macroinclusions during the hot processing of the steel (forging of ingots, rolling of bars and plates, extrusion of shapes).

A complete specification of a stainless steel for vacuum applications must consider the risk of leaks due to the presence of non-metallic micro- and macro-inclusions embedded in the microstructure of the final product. Segregations occurring during primary solidification are also pernicious. Indeed, the direct transformations of the primary melting ingots without a remelting process, such as ESR or Vacuum Arc Remelting (VAR), might not allow the ingot structure to be completely broken and homogenized and can result in unrecrystallized volumes and segregations in the final products [33].

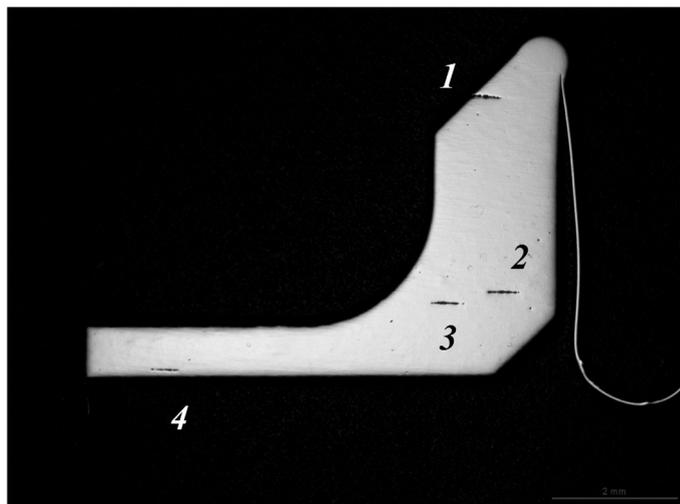

**Fig. 3:** Bellow end fitting machined from an AISI 316LN round bar, showing alignment of oversized (1,2,3) and thick (4) B type inclusions up to class 2, classified according to standard ASTM E45 [32]. CERN specification [23] imposes a class of inclusions at most 1 for type B inclusions and a half-class above this limit in up to 2% of the fields counted.

The risk of leaks can be reduced by one or more of the following actions.

1) In the technical specification, impose a maximum allowed content of non-metallic inclusions. CERN specifications impose maximum limits according to the standards in force.
2) Since microscopical test methods for determining the inclusion content of steel are not intended for assessing exogenous macroinclusions, impose a steelmaking process. Stainless steels are primary melted in an electric furnace and decarburized through an Argon-Oxygen Decarburization (AOD) or Vacuum-Oxygen Decarburization (VOD). These processes, consisting of a single melting step, might not guarantee alone a sufficient homogeneity of the ingot. Imposing a remelting process such as VAR or ESR ensures the homogeneity of the material will be effectively influenced. In addition, remelting reduces the impurity and the microinclusion content of the final products, allows high density and a lack of macrosegregations with shrinkage cavities to be achieved [34].
3) Specify three-dimensionally, redundantly forged products [24]. Upsetting steps allow the alignment of the inclusions which are elongated by two-dimensional steps of previous forging or rolling to be broken, thus reducing the risk of leak through the thickness.
4) Specify products with a fine and homogenous grain size.
5) Impose a final separate solution annealing rather than allowing mill-annealing from a high temperature process, in order to facilitate the achievement of a controlled, homogeneous and non-segregated microstructure [33], more easily inspectable by Non-Destructive Testing (NDT), see point 7 hereafter.
6) Avoid breaking the fibre of the product by machining walls perpendicular to the direction of the material flow, such as in the example of Fig. 3.
7) Introduce NDT such as Ultrasonic Testing (UT) not only on the final product, but also on semifinished products resulting from intermediate steps of steel processing. NDT procedures and acceptance criteria should be the object of an agreement between supplier and customer.

In conclusion, for vacuum applications to accelerators, stainless steels should never be selected on the basis of a mere material designation or availability of general purpose stock. Critical parameters should be explicitly specified and stringently controlled, in some cases not only on final product but through the definition and application of a quality control plan with tests associated with the different phases of material production.

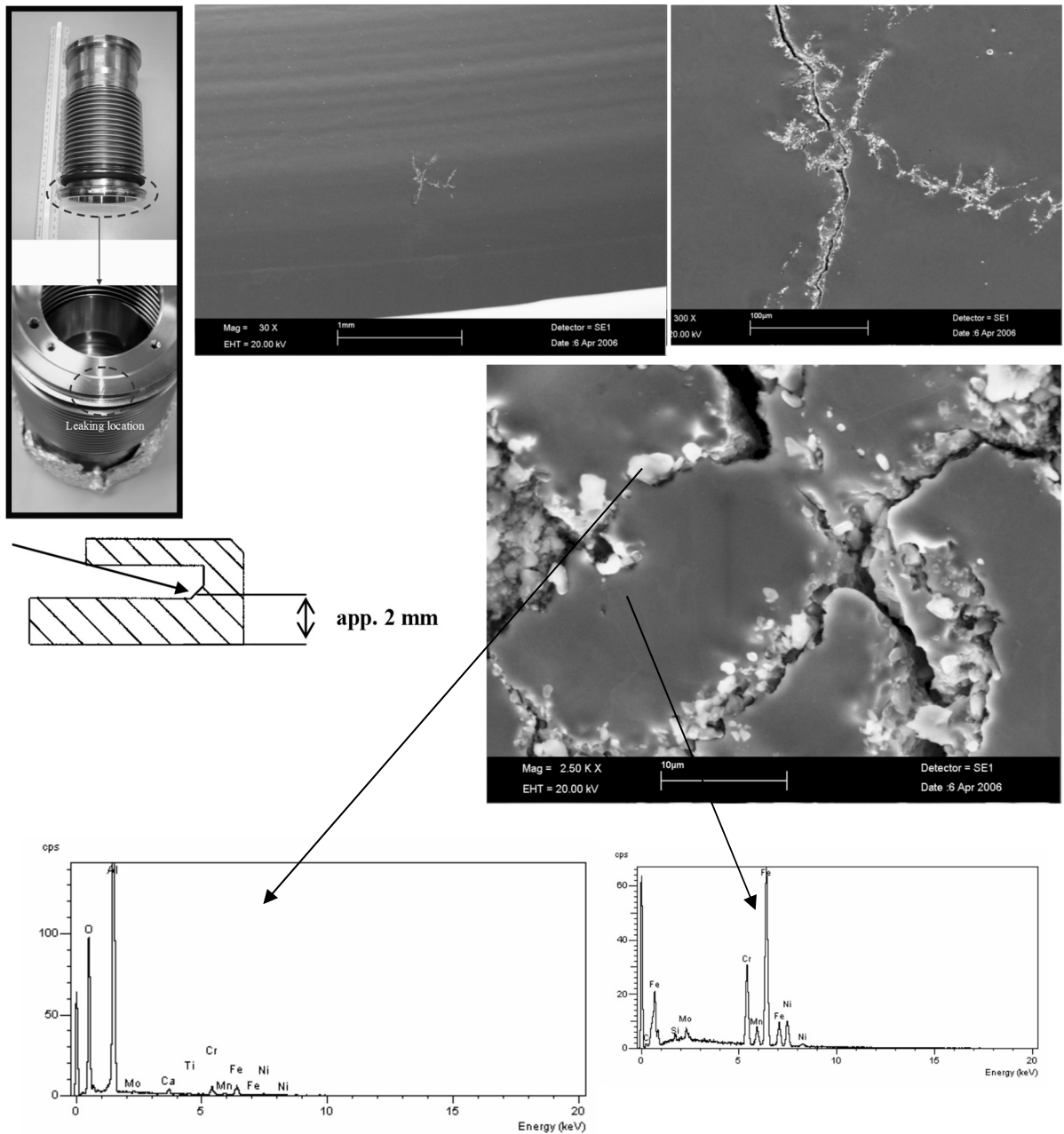

**Fig. 4:** Several plug-in modules of the LHC interconnections have been found leaking at the flange location. Flaws associated with the presence of Al, O, and Ca were detected on the two opposite surfaces of the throats of the cooling exit tube. Al and Ca are typical elements contained in slag or refractories. Entrapped residues of slag resulted in macroinclusions. Elsewhere, cross-sectional micro-optical observations of a non-leaking area only showed B type inclusion up to class 1 (worst field) [32].

## 2.5 Aspects related to joining processes of stainless steels in the framework of vacuum applications

### 2.5.1 Precautions during welding

Austenitic stainless steels are readily welded by conventional arc techniques such as Tungsten Inert Gas (TIG), MIG, or beam techniques such as laser and EB. Properly specified and qualified welds according to standards in force[1] can be fully sound. Tensile strengths equal to or higher than the minimum specified for the base metal can generally be obtained, as well as satisfactory fatigue crack growth rates, however for heavy gauge welds and/or specific combinations of parent and filler metals it might be challenging to obtain satisfactory joint strength [40]. Moreover, some additional aspects are relevant for welding austenitic stainless steels in the framework of a vacuum application.

#### 2.5.1.1 Possible presence of δ-ferrite in austenitic stainless steel welds

As mentioned in Section 2.2, austenitic stainless steels can contain residual amounts of δ-ferrite, which can be critical in applications to accelerators due to its reduced toughness and ferromagnetic nature. Constitution diagrams for stainless steel weld metals such as the one presented in Fig. 5 allow the ferrite content to be estimated in the as-deposited weld on the basis of the composition of the base and, when applicable, of the weld filler metal. Control of ferrite to minimum or zero level might be required in the context of specific accelerator applications (cryogenics, components close to the beam). Since fully austenitic stainless steel welds are more sensitive to microfissuring, specific precautions have to be taken such as reducing the weld heat input, minimizing restraint, designing for low constraint, and keeping impurities in the stainless steel composition to minimum levels [15].

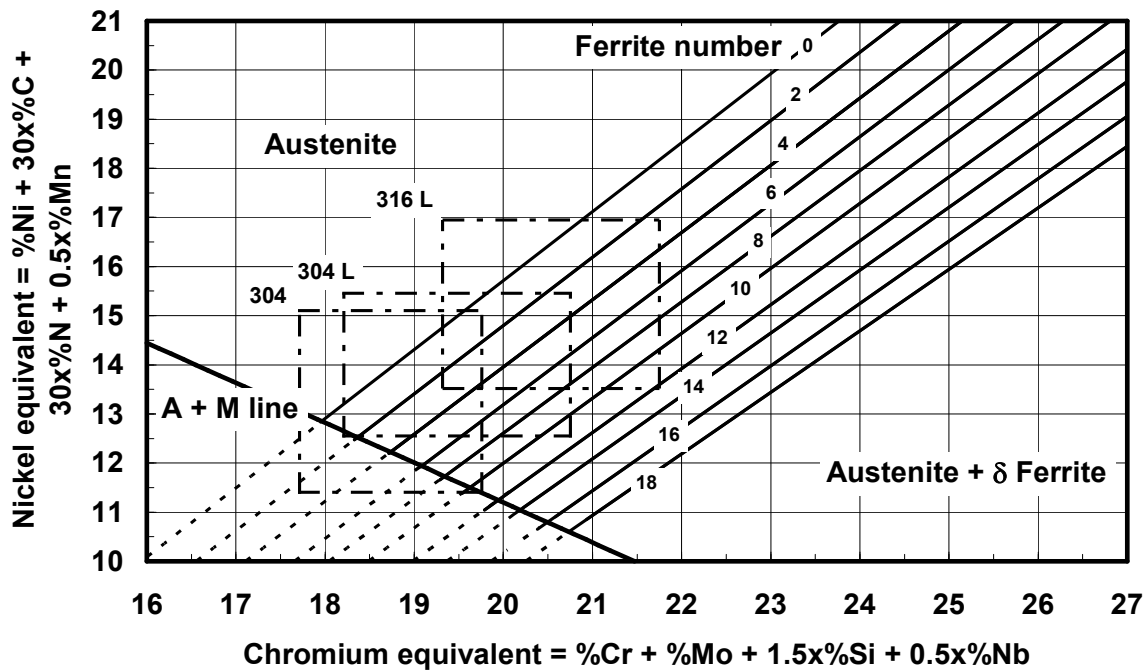

**Fig. 5:** Diagram for the determination of the ferrite content in austenitic stainless steel weld metal according to DeLong [41,42]. Diagrams more adapted to high N stainless steels are the ones of Hull [43] and Espy [44].

---

[1] Welds qualified according to standards ISO 15614-1 [35] for arc welding, ISO 15614-11 [36] for electron and laser beam welding, with reference to standard ISO 6520-1 [37], and level B of ISO 5817 [38] or ISO 13919-1 [39] for classification and quality levels for imperfections, respectively, are generally specified at CERN for vacuum applications.

*2.5.1.2 Minimizing the risk of microfissuring in fully austenitic stainless steel welds*

Figure 6 shows a so-called 'Suutala diagram' allowing the risk of hot cracking to be predicted, as a function of the solidification mode (primary ferrite or austenite) and the impurity content of the steel (P, S). The primary mode of solidification, be it austenite or ferrite, is important for predicting the integrity of the weld. The ratio of $Cr_{eq}$ to $Ni_{eq}$ identifies the two modes in the diagram, primary ferrite for $Cr_{eq}/Ni_e$ > 1.5 approx., primary austenite for $Cr_{eq}/Ni_e$ < 1.5. In the domain of solidification to primary austenite, only the grades, fillers or combination of them corresponding to a very limited total residual element content can be considered safe in terms of hot cracking. Particular care should be taken when mixing base materials of very different origin and quality. Austenitic stainless steels such as 316LN properly specified [23−25] generally solidify in the fully austenitic range and show limited impurity content. If mixed with a general purpose stainless steel such as 304L or 316L, or so called 'free machining' grades with added S to improve machining rates, a risk of cracking cannot be excluded depending on the dilution. Austenitic stainless steels with high impurity contents are generally designed to solidify in the primary δ-ferrite range, in order to avoid risks of hot cracking. Their dilution with a fully austenitic grade might locally result in an austenitic weld with equivalent unacceptable level of impurities.

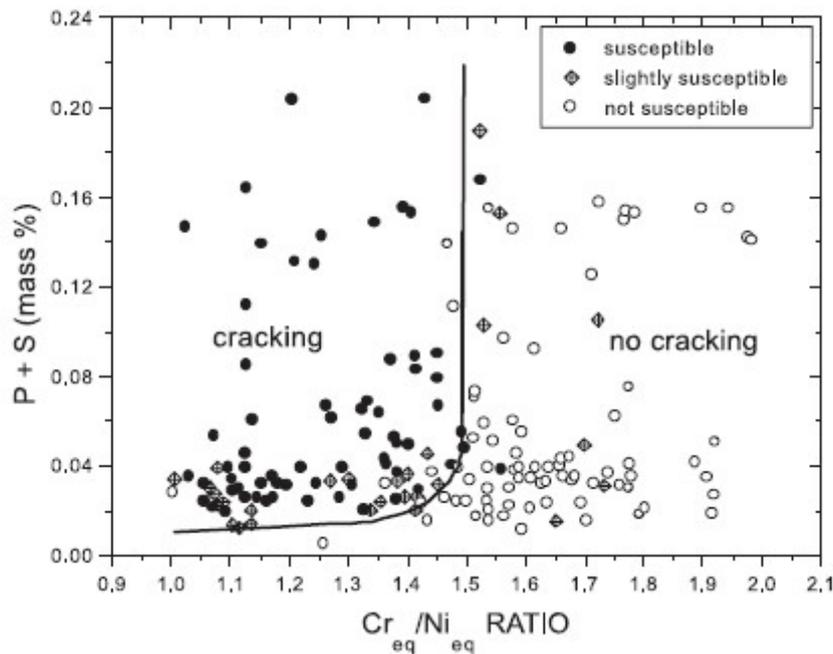

**Fig. 6:** Cracking susceptibility during arc welding of austenitic stainless steels ('Suutala' type diagram [45−48]). Equivalents based on Schaeffler equivalent formulae for $Cr_{eq}$ and $Ni_{eq}$, $Cr_{eq}$ = Cr + 1.5Si + 1.37Mo, $Ni_{eq}$ = Ni + 0.31Mn + 22C + 14.2N. Variants of the diagram exist to take also into account the effect of B.

Due to this risk, the use of free-machining grades, where addition of 150 ppm to 300 ppm S is allowed [26], should be strictly avoided. Components to be welded together in the vacuum system of an accelerator can be of very different origin and supplies. Their impurity content and solidification mode should be carefully checked to avoid leaks in the welds due to microfissuring.

*2.5.2 Solution annealing, outgassing treatments (vacuum firing), stress relieving*

Austenitic stainless steels are to be supplied and preferentially used in their solution annealed condition. Solution annealing consists of a heat treatment in a suitable temperature and time range adapted to the specific grade and size of the product, followed by quenching in water or rapid cooling by other means. Solution annealing allows optimum ductility and formability, toughness, and corrosion resistance to be achieved. Consistently, standards in force generally require the supply of products in the solution annealed condition, except for specific applications. Solution annealing is performed according to annealing requirements imposed by standards, at temperatures generally above 1040°C to avoid sensitization and precipitation of secondary phases. Solution annealing also allows ductility of cold finished products such as cold rolled sheets or drawn tubes to be recovered and hardness to be reduced. The maximum allowed hardness of products is limited by corrosion standards for use in severe environments [49]. CERN specifications also limit maximum allowed hardness.

Post-weld heat treatments are generally not required for austenitic stainless steels. On the other hand, for vacuum applications, a so-called 'vacuum firing' of components and subassemblies might be needed to outgas the material, in order to "remove the dissolved gas load in cleaned and degreased parts" [9]. 316LN products purchased to CERN specifications are compatible with applications requiring vacuum firing at 950°C. Even if 950°C is substantially below the solution annealing temperatures for austenitic stainless steels, a short treatment for a few hours at this temperature has no measurable detrimental metallurgical effects, owing to the presence of N in the grade that delays sensitization.

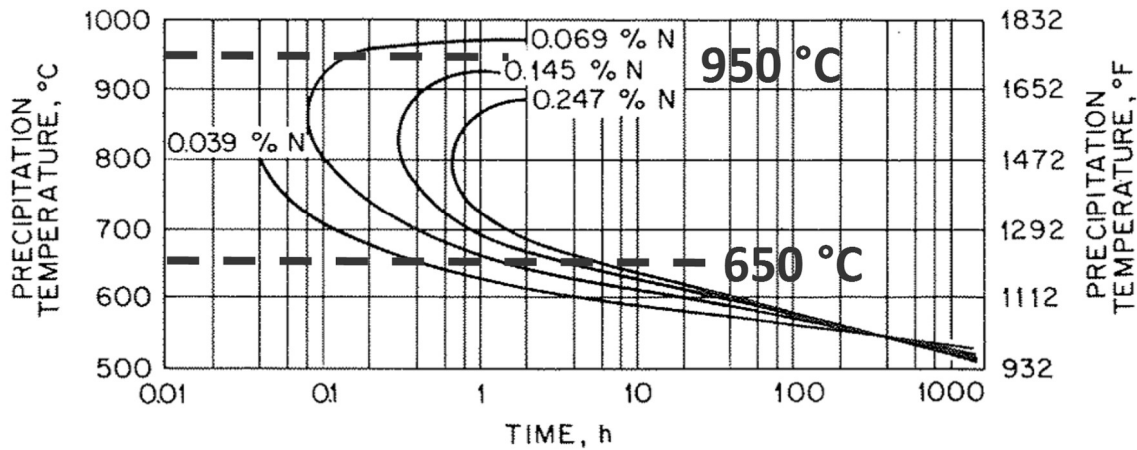

**Fig. 7:** Effect of N on precipitation of carbides ($M_{23}C_6$) in a 317 type stainless steel (0.05%C, 17%Cr, 13%Ni, 5%Mo) whose composition is close to 316LN. The N content range specified by CERN for 316LN is 0.14 % to 0.20 %. For such content, a treatment at 950°C in the few hour range is expected to have no metallurgical effects, while treating at 650°C for e.g., 24 h might provoke intergranular precipitation of carbides and sensitization (from Refs. [14] and [50]).

On the other hand, products in 316L (a grade nominally not destined to be vacuum fired) might also be submitted to vacuum firing for specific applications. Moreover, for 316LN products destined to be coated, vacuum firing at 950°C might be detrimental in function of the B content of the grade, and alternative treatments at 650°C in the tens of hour range are applied[2]. These treatments might result in metallurgical effects. Figure 7 shows the effect of holding time and temperature on a N bearing stainless steel (composition close to EN 1.4439 or 317LN, a grade similar to 316LN), as a function of the N content. For a standard vacuum firing at 950°C, no intergranular precipitation of carbides ($M_{23}C_6$) is expected for a few hour holding. This conclusion takes into account the specified N ranges of CERN (0.14% to 0.20% [24]) or standard (0.12% to 0.22% [26]) 316LN products and the low C content of 316LN (max. 0.030%). On the other hand, a treatment at 650°C for prolonged times at temperature induces metallurgical effects that should be carefully assessed.

---

[2] For a B content in solid solution above 8 ppm, vacuum firing of 316LN carried out at 950°C induces formation of BN at the surface [51], which is detrimental for the adherence of coated thin films and might require an eventual electropolishing prior to coating.

Figure 8 shows the detrimental effect of a heat treatment at 650°C and 200 h (to mimic the reaction heat treatment of Nb$_3$Sn) on the ductility at cryogenic temperature of a jacket specimen of the ITER Toroidal Field (TF) conductor in modified 316LN IG (ITER grade). When tensile tested at 7 K following a light cold work (compaction and 2.5% stretching to simulate bending and straightening as undergone by the jacket during coil winding) and the above heat treatment, the specimen featured a ductility of only 11%, in spite of a severe specification limiting the C content to 0.02% max. (with a target < 0.015% against sensitization). A tight control of chemical composition and manufacturing processes eventually allowed an elongation at breakdown of at least 20% to be maintained [52]. For comparison, standard 316LN grade in the solution annealed condition can feature ductility at 4.2 K above 40%.

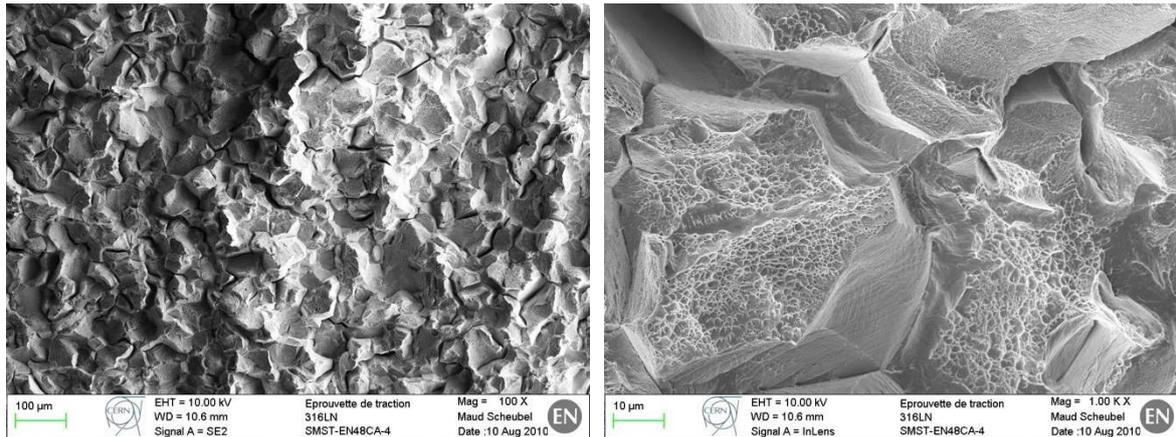

**Fig. 8:** Scanning Electron Microscope (SEM) fractographic analysis of a compacted and aged 316LN IG specimen of the jacket of the ITER toroidal field (TF) conductor tensile tested at 7 K, having being sensitized by a 650°C – 200 h treatment following a light cold working. The observations confirmed quasi-cleavage with mainly intergranular fracture at the grain boundaries and limited dimpled areas within the single grains [53].

Stress relieving should also be preferentially operated outside the sensitizing temperature range, either at low temperatures (i.e., under 500°C for 316LN) or at 950°C or above, where it can be made coincident with the 950°C vacuum firing treatment. Additional restrictions might apply to welded structures, where σ-phase precipitation inducing detrimental effect on toughness can occur even above 950 °C. This is particular relevant for Mo bearing grades and if the welds contain δ-ferrite, depending upon the ferrite content of the weld metal [42].

Sensitization implies loss of corrosion resistance (Cr depletion at grain boundaries), loss of ductility (especially at cryogenic temperatures) with a ductile-to-brittle transition onset. The effect of a thermal treatment and the possible occurrence of sensitization in austenitic stainless steels can be checked with the support of ASTM A262 standard, in particular through practice A and E [54].

## 3   Aluminium alloys

Wrought aluminium alloys are very attractive materials for ultrahigh vacuum systems for accelerators. Aluminium alloys allow the residual radioactivity after machine shutdown to be strongly reduced compared to stainless steels, show high transparency to radiation, can be shaped into complicated profiles by extrusion and drawing, are completely non-magnetic, and, for given components, might be competitive in terms of cost. High electrical and thermal conductivity are an asset. High thermal conductivity and low thermal emissivity allow aluminium alloy components to tolerate high heat fluxes in spite of their relatively low melting point [55]. An all-aluminium alloy vacuum system has been adopted in the frame of the TRISTAN electron-positron collider constructed at the National Laboratory for High Energy Physics in Japan, following developments of components such as bakable aluminium vacuum chambers and bellows [55,56].

## 3.1 Aluminium alloy grades of interest for vacuum construction and their weldability

Aluminium alloys are classified in eight groups or series (Table 2), according to their main alloy element. The first digit identifies the family, the last two digits the type in the family (with the exception of the 1xxx series where subsequent digits identify the purity), the second digit a variation in the same family.

**Table 2:** Wrought aluminium alloy designation and grades of practical interest for vacuum applications. Heat treatable families are shown in bold characters. Some alloys of the 8xxx series are heat treatable as well.

| Alloy group | Designation AA | Readily weldable alloys of the family | Examples of relevant alloys for vacuum applications |
|---|---|---|---|
| Pure aluminium | 1xxx series | EN AW-1060, -1100, -1199, -1350 | |
| **Al-Cu** | **2xxx series** | **EN AW-2219, -2090 (AlCuLi), -2050 (3$^{rd}$ generation AlCuLi alloy)$^3$** | **EN AW-2219** |
| Al-Mn | 3xxx series | EN AW-3003, -3004, -3105 | EN AW-3003 |
| Al-Si | 4xxx series | weld fillers | weld fillers |
| Al-Mg | 5xxx series | EN AW-5005, -5050, -5052, -5083, -5086, -5154… | EN AW-5083 |
| **Al-Mg-Si** | **6xxx series** | **EN AW-6061, -6063, -6070, -6082…** | **EN AW-6082 (-6061)** |
| **Al-Zn** | **7xxx series** | **EN AW-7005, -7020** | |
| Al+other element (e.g., Li) | 8xxx series | **EN AW-8090 (AlLiCu)** | |

Aluminium alloys can be hardened by cold or warm working (strain hardened alloys) and/or by heat treatment (age hardened alloys), depending on the type of alloying elements. For strain hardened alloys, the level of mechanical properties that can be attained depends on the alloying elements and their content: 5xxx alloys have a potential level of mechanical strength achievable by work hardening that is superior to the 1xxx and 3xxx series. Age hardening alloys, identified in bold in Table 2, are generally strengthened by a three stage treatment (solution annealing, quenching or rapid cooling, holding at RT or higher temperature). A cold work step can be added between solution heat treating and aging for certain alloys.

Extrudable (EN AW-6061, -6082, -5083, 7020…) and weldable alloys (see Table 2) are of paramount importance for vacuum constructions for accelerators. Several alloys within the different families are considered as readily weldable. The weldability of aluminium alloys by EB techniques is rated by the Merkblatt DVS 3204 [57]. Weldability of Li-bearing aluminium alloys is discussed in Ref. [58], while the more recent third generation AlCuLi alloy 2050 is described in Ref. [59].

---

$^3$ Mainly by friction stir welding.

Fine grained alloys can be produced via a melt spinning route, a rapid quenching process allowing cooling rates up to $10^6$ °C/s. Following melt spinning and chopping, the flakes are consolidated into a compacted billet that undergoes a conventional extrusion process [60]. A very fine microstructure results from this process, associated to high mechanical properties. Proprietary alloys with a composition Al-Mg5-Mn1-Sc0.8-Zr0.4, Al-Fe2.5-Ni5-Cu2.5-Mn1-Mo0.8-Zr0.8, Al-Fe8.7-Si1.8-V1.3 have been recently studied at CERN, achieving a yield strength in the range 385–510 MPa in function of the alloy [61].

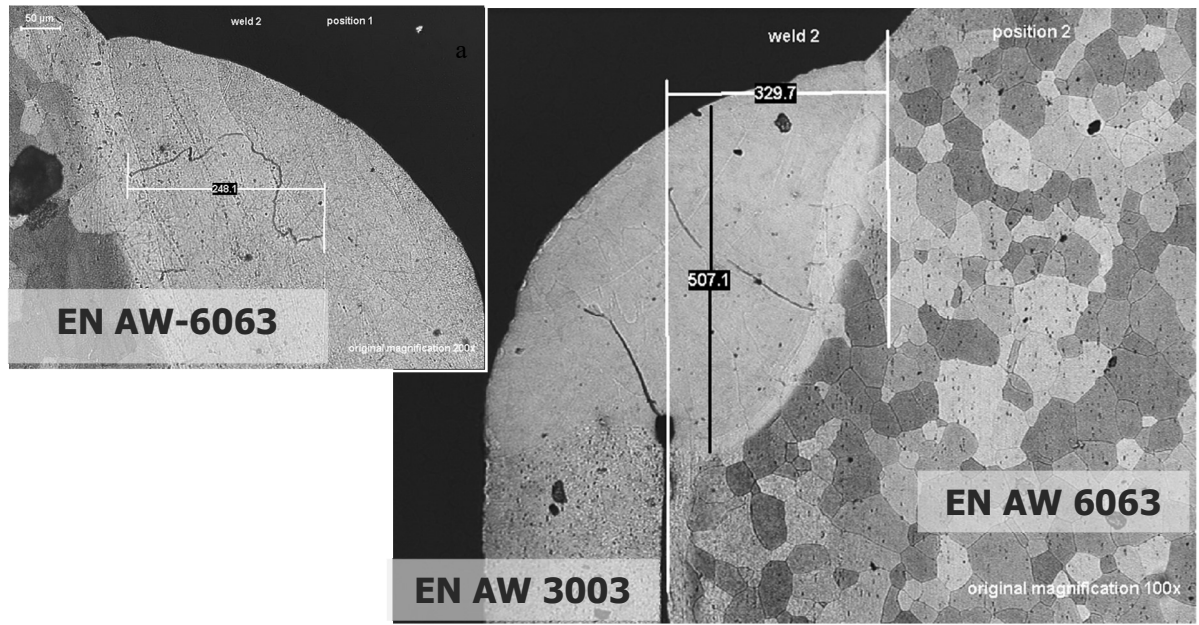

**Fig. 9:** Solidification cracks in autogeneous laser weld of tubes and fittings of the cooling circuit of the ATLAS Pixel experiment involving a 6xxx series alloy: a) laser weld developing a liquation crack; b) in mixed welds (EN AW-6063 diluted with EN AW-3003), where no filler is applicable (laser weld of thin walled components), cracks can develop due to an unfavourable dilution of the two alloys. Locally in the weld bead a critical Mg-Si balance can be present, resulting in solidification cracks [62]. Measured distances in μm.

Some families of weldable alloys are more sensitive than others to microcracking (solidification cracking). In general, the Si and Mg contents of extrudable alloys of the 6xxx series (Al-Mg-Si) are optimized for extrusion purposes. Si and Mg build the intermetallic phase $Mg_2Si$ that is soluble to 1.85% at the eutectic temperature. Solidification cracking (Fig. 9) appears when solidifying weld metal undergoes tensile stresses during its solidification [63].

Whenever applicable (typically in arc welding of products above few mm thicknesses) the use of a filler metal can reduce the risk of weld cracking. Detailed tables are available suggesting the most suitable fillers for different alloys and their combinations [64]. These tables rate the fillers as a function of ease of welding, strength of the joint, ductility, corrosion resistance, service at moderate or high temperatures…

In addition to the risk of cracking, porosity is a common problem when welding aluminium alloys. Porosity arises from gas entrapment due to poor shielding, air, moisture, unclean wire or metal surface (hydrated oxides, oil, hydrocarbon contaminants), high cooling rates… Hydrogen contamination is the cause of virtually all weld porosity in aluminium alloys, due to its high solubility in the molten pool and limited solubility in solid aluminium.

For the specific case of pure Al, the role of minor elements, especially Na and Ca, has been identified as cause of welding issues (cracks, voids, and excessive porosity). The weld response of different pure Al heats containing different levels of minor elements [65] was compared. The role of Na in the crack sensitivity of Al alloys was already identified by Ransley and Talbot [66]. Na is sometimes added to outgas pure aluminium in the melting phase [67]; Ca has a known interaction with hydrogen [68].

## 3.2 Compatibility of aluminium alloys with high temperature service (baking, activation of Non-Evaporable Getters)

Non heat-treatable alloys, such as EN AW-5083, one of the most common general purpose Al-alloys, can be supplied in a work hardened state, where they can reach significant strength. The same alloys in a fully annealed or slightly cold strained temper (designed as O and H111, respectively), show moderate to low tensile properties which might be of limited interest in several applications.

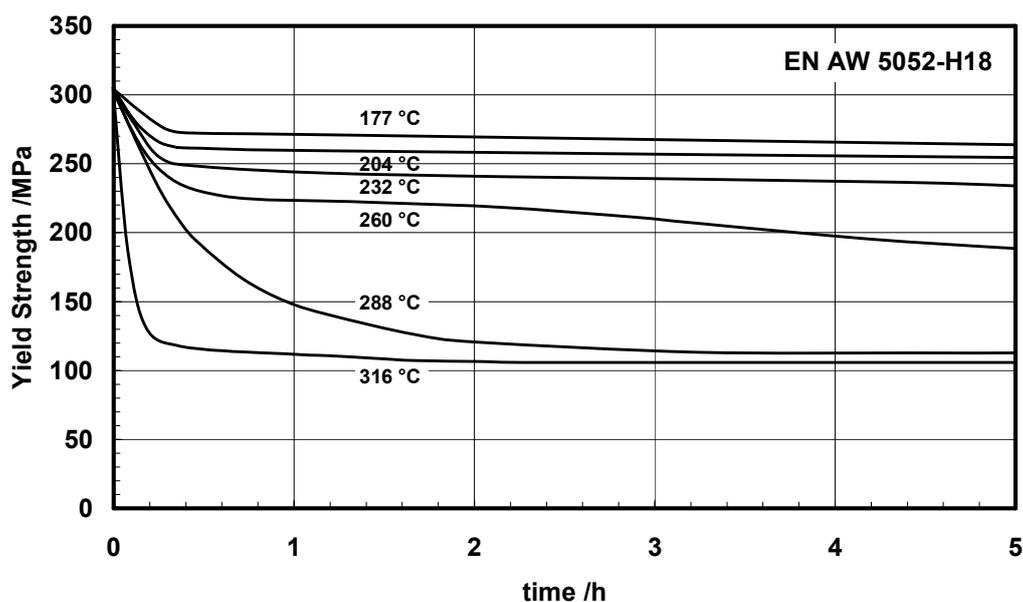

**Fig. 10:** Annealing curves of work hardened EN AW-5052 [69]. At a given temperature (curves for 177°C ≤ T ≤ 316°C are reported), increasing time at temperature results in a significant loss of the initial yield strength. The reported annealing temperature of this alloy is 343°C [69], corresponding to the temperature allowing for full annealing with no time at temperature required.

In the example of Fig. 10, the annealing curves of a typical wiredrawing alloy such as EN AW-5052 are presented. In a fully hard state designed as H18 temper [70] and achieved by about 75% cold reduction and temperature control during reduction (not to exceed 50°C), EN AW-5052 can reach a yield strength above 300 MPa. The alloy is softened by the exposure at temperature for times of few hours.

If a time-temperature cycle has to be applied to an aluminium vacuum chamber for baking purposes or for activation of Non-Evaporable Getters (NEG), the possible loss of properties induced by the cycle should be critically considered. Let us take as an example a cumulative effect of 30 activations over 24 h at 260°C. One day cycles of activation at this temperature per year are applicable to TiZrV NEG coatings deposited on the internal walls of vacuum chambers for the intersection areas of LHC. In the example shown in Fig. 10, a sudden loss of properties would occur for EN AW-5052 H18 already after the first hours of application of the heat cycle. The final properties will tend to the ones of a fully annealed state.

In addition to the properties at RT after exposure to high temperatures, the properties at the temperature of activation or baking are relevant for the design of a component to be exposed to high temperature cycles. The total time of exposure at temperature should be taken into account. Creep considerations should not be forgotten in design, especially when dimensioning a chamber to resist buckling [71].

Heat treatable alloys show a different response to a thermal cycle. In the example of Fig. 11, the heat treatable alloy EN AW-6061 is initially in a solution annealed and naturally (at RT) aged temper. Temperature cycles result in an initial increase of strength up to a maximum followed by a further decrease (overaging). At $T = 260°C$, a few days of exposure is sufficient to reduce the strength to very poor values. The yield strength expected after 1000 h at 260°C is only 70 MPa [72], far below the initial 275 MPa of the product delivered in a peak-aged temper.

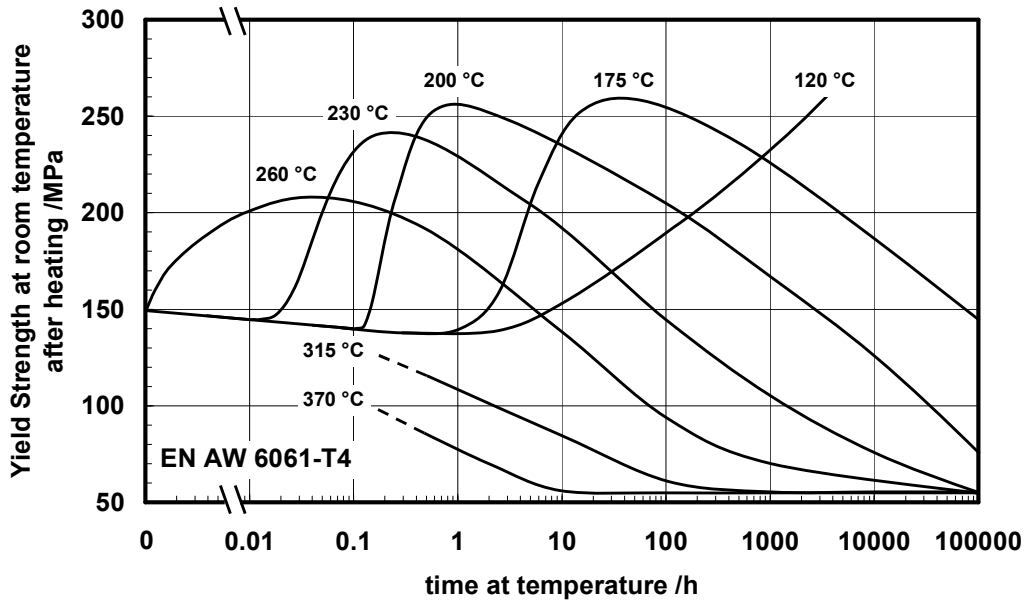

**Fig. 11:** Effect of time and temperature on the yield strength of EN AW-6061 [69]

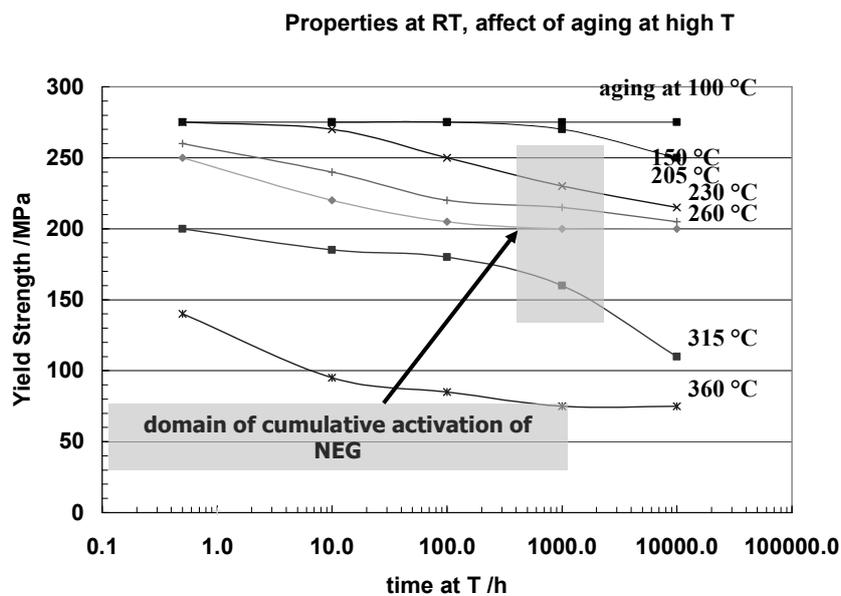

**Fig. 12:** Effect of time and temperature on the yield strength of EN AW-2219 (from data of Kaufman [72]). Activation of TiZrV NEG coatings typically occurs in a temperature range between 200°C and 300°C.

In summary, considerations of compatibility with heat treatment cycles severely restrict the selection of Al-alloys for vacuum applications. The effect should be critically discussed in the phase of alloy selection. A weldable high strength alloy, EN AW-2219, developed for high temperature application by Alcoa in the 1960s, is a solution for applications where the loss of mechanical properties at high temperature is a concern. As from the plot of Fig. 12, starting from a proper temper state, this alloy maintains a yield strength of approximately 200 MPa after 1000 h at 260°C.

### 3.3 Thin walled components: a failure analysis

EN AW-2219 bellows were machined from a forged rod. These bellows, to be integrated in one welded assembly (two flanges + two bellows + one tube), were intended for application to LHCb, one of the four major experiments of LHC. One bellow revealed a leak, detected on a convolution and issued from the presence of corrosion pits on the inner surface of the convolution. SEM observations and Energy-Dispersive X-ray spectroscopy (EDX) analyses were performed at the pitting corrosion location. They revealed the presence, among other elements, of Cl [73]. Cross-sectional observations (Fig. 13) confirmed the intragranular nature of the pits and showed a rough grain microstructure (Fig. 14), consisting of one or few grains in the wall thickness of the convolutions. Presence of Cl residues issued from the machining operations, performed with the help of Cerrobend contaminated by halogens from previous machining residues, was sufficient to provoke through thickness leaks [74].

Hydraulic forming of seamless tubes to produce Al-alloy bellows 0.3 mm thick is also reported in the literature, as well as fabrication of bellows by welding and brazing. CERN has recently developed and successfully applied a route for the fabrication of thin walled aluminium alloys bellows by warm hydroforming starting from rolled and welded EN AW-5083 H111 tubes [75]. There is an intention to apply the same method to seamless drawn thin walls to be specially produced in the same alloy.

In case of thin walled products to be machined from bulk material such as bellows and windows, special three-dimensional forgings should be ordered with a very fine grain size. As an example, starting material for the window of the VErtex LOcator (VELO) of the same LHCb experiment, including a thin walled window and bellow, consisted of grains of a few micrometres in size. This microstructure could be achieved through free forging of heavy blocks of EN AW-6061 at controlled temperature. NDT of the semifinished products at different stages of transformation was also applied to detect possible discontinuities at an early stage. Immersion or mechanized contact UT techniques allow submillimetre defects to be identified in due time. The price of the special forgings for the VELO window was 12.5 EUR/kg, to be compared with the price of 5 EUR/kg of standard plates in 5xxx or 6xxx alloys.

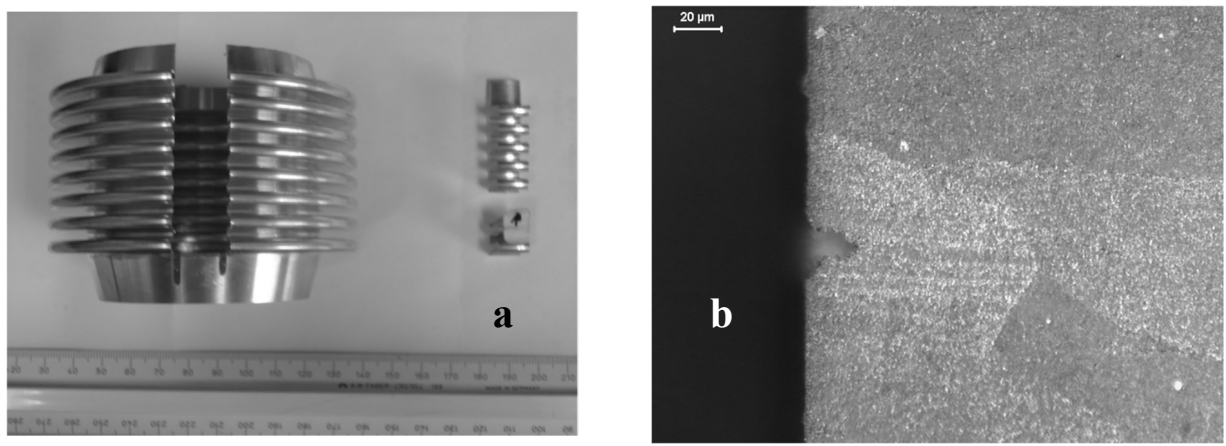

**Fig. 13:** Thin walled (0.4 mm) bellow machined from a generic purpose forged rod. (a) A sample is removed from the bellow at the location of the leak; (b) cross-sectional observations.

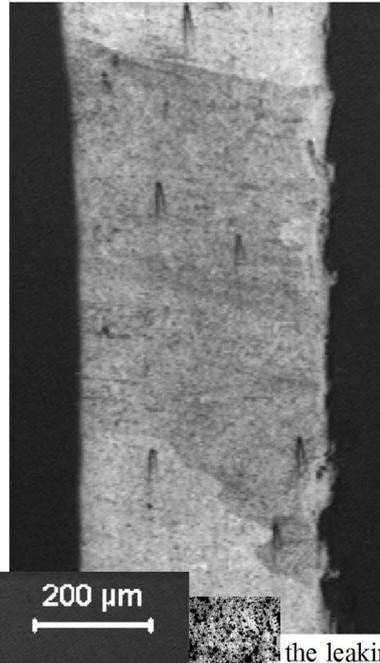

**Fig. 14:** Comparison at the same scale [of] the leaking bellow, showing one or a few grains through the wall thickness of the convolution, and the fine grained microstructure of a specially developed Al-alloy forging for the VELO window of the same LHCb experiment.

# 4 Copper and copper alloys

A relatively high corrosion resistance, machinability, and formability into different shapes, as well as the ease of weld for several grades make copper and copper alloys very attractive for vacuum components, especially for those applications requiring high electrical and/or thermal conductivity. Table 3 lists the distinct families of wrought copper and copper alloys according to their composition. Standard designation systems for coppers and copper alloys include the Unified Numbering System (UNS) [76], and a temper designation system [77].

Relevant for vacuum applications are pure coppers and the so called 'electrical coppers', belonging to the first group in Table 3 (coppers, C10100-C15815), high strength copper alloys such as high and low Be Cu-Be alloys (C17200 and C17410, respectively), Cu-Ni alloys such as the non-magnetic '70-30 cupronickel' (C71500). On the other hand, alloys containing Zn, Pb, Cd, Se, S… might result in unsuitable vapour pressure for vacuum applications [8].

**Table 3:** Classification of wrought coppers and copper alloys (from [78])

| Wrought alloys | UNS No. | Composition |
| --- | --- | --- |
| Coppers | C10100-C15815 | >99 % Cu |
| High-copper alloys | C16200-C19900 | >96 % Cu |
| Brasses | C20100-C28000 | Cu-Zn |
| Leaded brasses | C31200-C38500 | Cu-Zn-Pb |
| Tin brasses | C40400-C48600 | Cu-Zn-Sn-Pb |
| Phosphor bronzes | C50100-C52480 | Cu-Sn-P |
| Leaded phosphor bronzes | C53400-C54400 | Cu-Sn-Pb-P |
| Copper-phosphorus and copper-silver-phosphorus alloys | C55180-C55284 | Cu-P-Ag |
| Aluminium bronzes | C60800-C64210 | Cu-Al-Ni-Fe-Si-Sn |
| Silicon bronzes | C64700-C66100 | Cu-Si-Sn |
| Other copper-zinc alloys | C66300-C69710 | Cu-Zn-Mn-Fe-Sn-Al-Si-Co |
| Copper nickels | C70100-C72950 | Cu-Ni-Fe |
| Nickel-silvers | C73500-C79830 | Cu-Ni-Zn |

## 4.1 Pure coppers

Oxygen-free copper (Cu OF, C10200, 99.95% min Cu) and Oxygen-free, electronic copper (Cu OFE, C10100, 99.99% minimum Cu) are high conductivity electrolytic coppers (in the annealed state, conductivity >101% IACS at 20°C) with limits for oxygen and other impurities. For Cu OFE, there are specific limits for 17 elements [79]. These grades are typically used for the fabrication of RF cavities, bus bars, waveguides, vacuum seals, klystrons, sputtering targets, evaporation materials for thin film applications, etc.

OFE grade is to be preferred for applications involving vacuum brazing or EB welding, and for cryogenic applications where a high Residual Resistivity Ratio (RRR) is required. Fine grained products are mandatory for vacuum applications, especially when thin walled components are foreseen. CERN specifies a uniform grain size and minimum allowed grain size number according to ASTM E112 of 4, i.e., a maximum average grain size of 90 μm [80]. The possibility of achieving a specified grain size not only depends on the sequence of the applied thermomechanical processing, but also on the grain structure and texture of the ingots supplied for further forging or drawing (see Fig. 15). The supplied forged bars should be 100% ultrasonically inspected to detect possible continuity faults. The agreement between the supplier and the customer of an UT procedure is essential to define a control of the material quality in the framework of critical vacuum applications. As an example, CERN specifies the highest possible control frequency, ideally 4 MHz, but this depends on the thickness. The homogeneity and the grain fineness of the bars are imposed through a criteria of 20% maximum allowed ultrasonic attenuation of the first back wall echo at any point in the product [80].

For ease of machining, quarter-hard tempers are preferred to fully soft conditions. OFE Cu at mid-2017 prices is supplied at a cost of 25−40 EUR/kg in 3D forged products. The base price of OF Cu is 10 EUR/kg. Other coppers such as Electrolytic tough-pitch (ETP) copper (99.90 % min Cu, Ag is counted as Cu), due to their oxygen content, are subject to embrittlement when heated at 370°C or above in a reducing atmosphere, and might be critical for applications involving annealing, brazing, or welding.

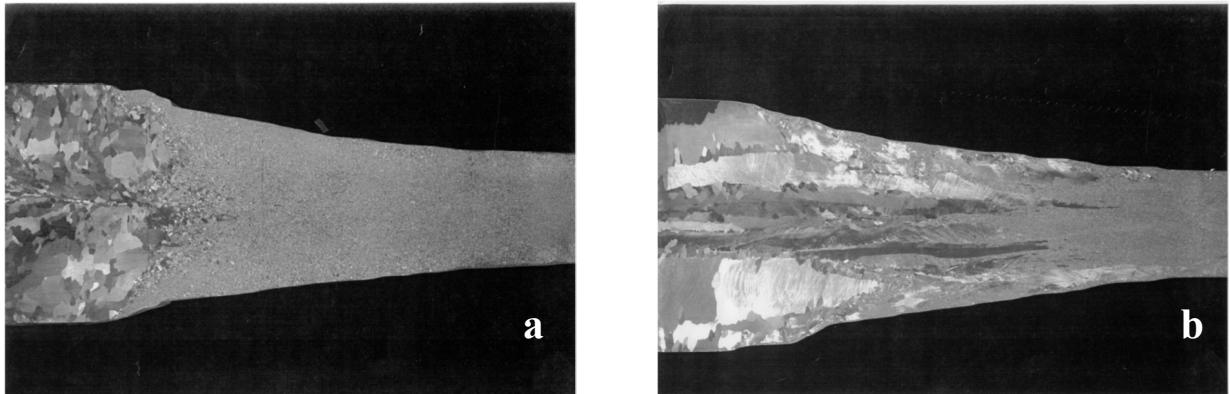

**Fig. 15**: Partially forged pure copper ingots from two different suppliers. Diameter of the incoming ingot, approx. 200 mm. The final grain size achieved by the same forging sequence applied to the two ingots strongly depends on the initial texture: (a) fine grained solidification structure, (b) rough grain structure (courtesy of Stainless).

## 4.2 Copper-silver and copper-zirconium alloys

As with many metals and alloys, electrical coppers, and in particular pure coppers, undergo thermal softening, a degradation of strength and hardness after exposure to elevated temperatures. Oxygen-free copper is available as silver-bearing copper having a specific minimum Ag content. Ag increases the resistance to softening, in particular in cold worked states, without substantially decreasing thermal and electrical conductivity at RT. Annealed oxygen-free silver coppers have an electric conductivity of 100% IACS at 20°C. Figure 16 shows the annealing curves of cold worked OF copper, and of different

OF coppers containing increasing amounts of Ag [81]. For the same amount of initial cold work and annealing time (1 h for the curves of Fig. 16), increasing amounts of Ag displace the annealing temperature toward high temperatures.

In addition, Ag improves the creep strength of pure copper. NEG coatings, deposited as thin magnetron sputtered TiZrV films on the internal wall of the several hundred drift chambers of the vacuum system of the long straight sections (LSS) of the LHC, are compatible with an activation in a temperature range between 200°C and 300°C (see Section 3.2). In view of the repeated activations, alloy C10700 (99.95% min Cu+Ag, 0.085% min Ag, Oxygen max: 0.0010%) has been selected for the chambers. A hard drawn temper with specified maximum mechanical property limits has been preferred in order to maintain substantial strength after exposure to thermal cycles. Since increasing initial work hardening has the effect of decreasing, ceteris paribus, the annealing temperature of the alloy, the maximum allowed initial hardness was limited to 90 HRF (Rockwell Hardness F Scale).

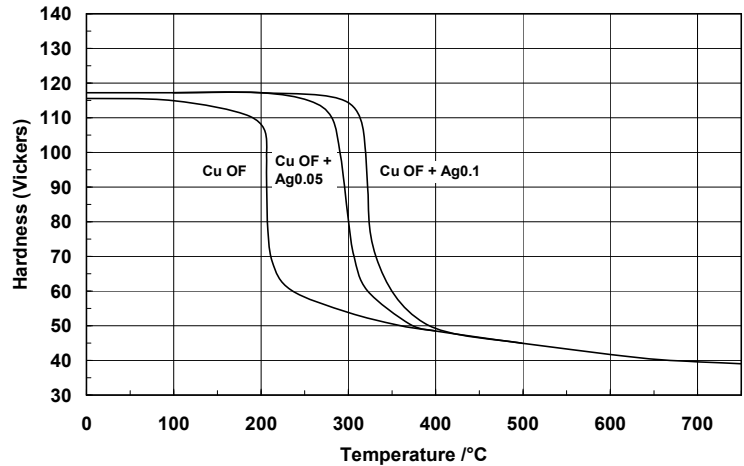

**Fig. 16:** Softening characteristics of cold worked OF Cu containing various amount of Ag. Annealing time is 1 h [81].

An alloying element that is also capable of increasing the annealing temperature of pure Cu is Zr. The resistivity increase of Zr per 1% addition is 80 n$\Omega$·m, much higher than Ag (3.6 n$\Omega$·m). A typical value of conductivity of CuZr alloys (0.10% $\leq$ Zr $\leq$ 0.20%) is 93−95 IACS and 76−90 IACS for CuCr1Zr. Zirconium coppers are heat treatable and their conductivity and strength depend strongly on the temper state. The alloys show improved fatigue strength and retain much of their room-temperature strength up to 450°C.

## 4.3   Oxide Dispersion Strengthened (ODS) coppers

Pure copper can be strengthened by a small amount of finely dispersed oxide, such as alumina, in the matrix. The dispersoids are stable at elevated temperatures, up to the melting point of the matrix, and prevent recrystallization and softening of the material when exposed to high temperatures. Examples of ODS coppers are GlidCop (trademark of North American Hoganas High Alloys LLC, based on a Cu-$Al_2O_3$ system) and CEP DISCUP (developed by CEP – Compound Extrusion Products GmbH).

Since the oxides are immiscible in liquid Cu, PM techniques are applied followed by a consolidation passing through conventional thermomechanical steps or HIP [78]. Several techniques are applied to achieve a fine and uniform dispersion of oxides, such as selective internal oxidation [82], mechanical mixing [83], and coprecipitation from salt solutions [84]. Starting from a copper-aluminium solid solution, aluminium is converted into alumina during the process.

Figure 17a [85] shows the effect of increasing aluminium oxide content on strength, hardness, ductility, and conductivity of ODS coppers. Figure 17b [86] shows the significant improvement of the resistance against high temperature softening conferred by the addition of 0.25% of Al in the form of $Al_2O_3$ to pure Cu. Compared to alloys based on additions of soluble elements, the improved resistance to recrystallization and softening is particularly relevant above 600°C (vacuum brazing temperature range). This is due to the inertness of the dispersoids compared to the easily dissolved precipitates in conventional precipitation hardened alloys such as CuZr or CuBe, based on diffusion controlled systems which are thermally unstable [87].

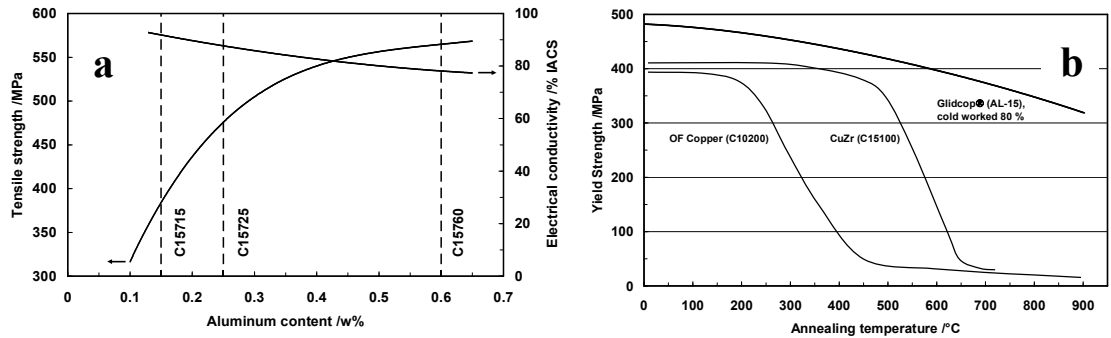

**Fig. 17:** (a) Properties of ODS coppers (rod stock in hot extruded condition) [85]. (b) Softening resistance of cold worked Glidcop AL-15 compared to OF copper and copper-zirconium. Glidcop AL-15 is reinforced with 0.15 % of Al as $Al_2O_3$ and roughly corresponds to grade C 15725 of Fig. 17a. Annealing time is 1 h [86].

Alumina reinforcement particles are of a typical size ranging from a few nm to a few tens of nm (Fig. 18a). Although of different shape, their size is comparable to that of the strengthening phases of the highest performing copper alloys such as CuBe (Fig. 18b).

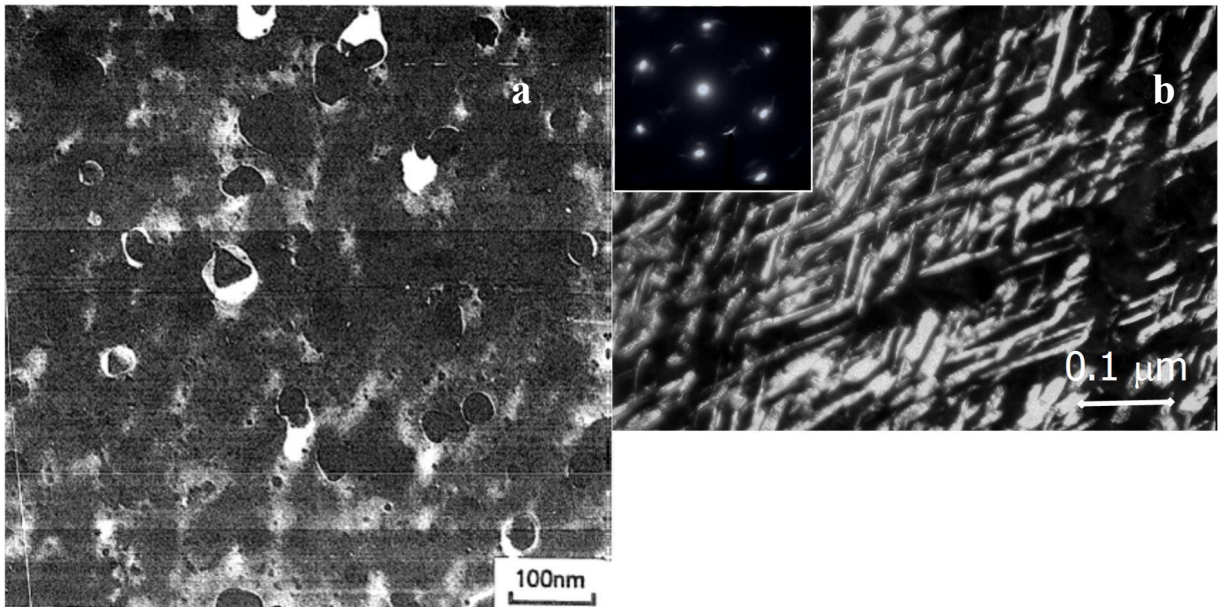

**Fig. 18:** (a) Alumina dispersed precipitates [88]. (b) At approximately the same scale, image of the reinforcing phase of a CuBe alloy C17670; dark field transmission electron microscopy (TEM) view of $\gamma'$ precipitates for a solution annealed and artificially aged (360°C – 6 h) temper [87].

GlidCop can be brazed with silver braze alloys, but requires a previous electroplating with Ni or Cu of the surfaces to be brazed. Gold-copper braze alloys of unplated Glidcop have been successfully used and are reported in the literature [86]. EB welding tests have been also performed at CERN with successful results [89]. Glidcop is largely used in accelerator components such as the crotches of the ESRF [90], exposed to intense bending magnet radiation, and in the collimators of the LHC. A typical current price for Glidcop is 55 EUR/kg.

# 5 Special technologies

## 5.1 Near-net shaping of the covers of the LHC dipole magnets [91]

The 1232 superconducting dipole magnets of LHC operate at 1.9 K. The cold mass of the dipole magnets is enclosed by a shrinking cylinder and two end covers at each extremity of the cylinder. The end covers are domed and equipped with a number of protruding nozzles for the passage of the different cryogenic lines (Fig. 19). The covers are structural components that must retain high strength and toughness at cryogenic temperature. They are welded by Metal Active Gas (MAG) onto the magnet shrinking cylinders. The protruding nozzles of the covers are welded to the interconnection pipes by an automatic orbital autogeneous TIG technique. Several thousand welds are necessary. AISI 316LN has been selected because of its mechanical properties, ductility, and the stability of the austenitic phase against low-temperature spontaneous martensitic transformation. Due to the complex geometry of the end covers, a Powder Metallurgy (PM) + HIP technique was selected for the fabrication of the covers. PM is an attractive near-net-shaping technique that allows the final shape to be approached and the machining to be reduced to a minimum. The covers were produced from atomized powders of the relevant steel grade, blended, homogenized, and poured into capsules with a geometry approaching the cover shape. After evacuation and sealing, a HIPing cycle was performed. HIPing consists of a time-temperature-pressure cycle performed at 1180°C for 3 h under a stress of 100 MPa, allowing a fully dense structure to be achieved. The capsulated covers were solution annealed, water quenched, pickled to remove the capsules, ground, and machined to the final dimensions. A 100% dye penetrant, visual, and UT inspection (to measure the wall thickness and detect possible defects) was finally performed on the finished covers, showing no relevant defects and full soundness and compactness of the components. Closed or open die forging would have required significantly more machining, a welded product would have needed extensive inspections and stress relieving, whilst a cast solution would have featured poorer mechanical properties. PM was demonstrated as a technique fully adapted to the fabrication of complex shape components, such as LHC end covers, working in a severe cryogenic environment. The covers, after several years of successful operation in the LHC, have proven to be perfectly leak tight to superfluid helium. This near-net shaping technique, retained for the series production, was also selected on the basis of its price competitiveness.

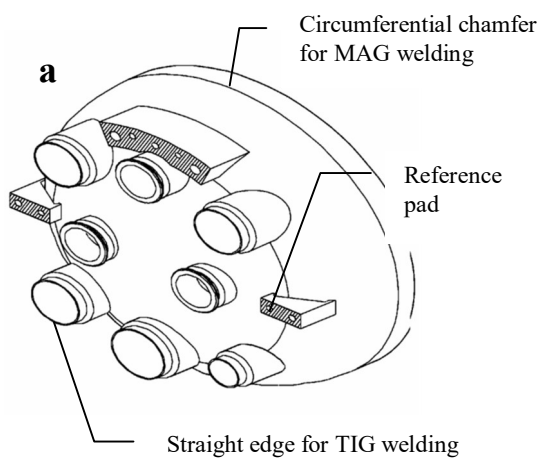
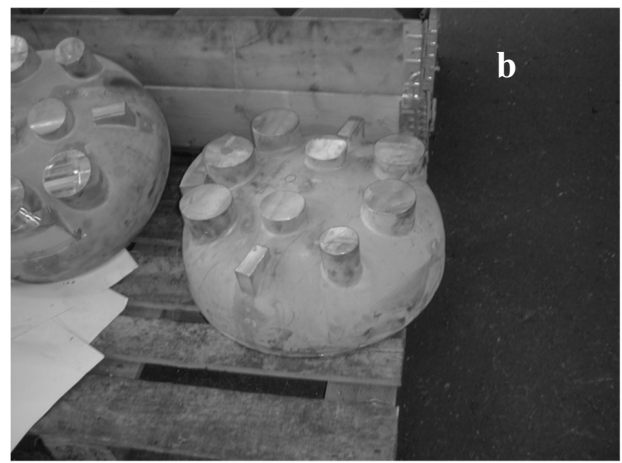

**Fig. 19**: (a) A 3D view of an end cover. The diameter of the cover is 570 mm, the height 224 mm, and the overall weight 100 kg. The extremities of the nozzle to be welded on the interconnection pipes have an external (internal) diameter of 84 (80) mm. The main circumference to be welded to the shrinking cylinder has a thickness of 10 mm. The most severe dimensional tolerances are of the order of 0.05 mm. (b) Near-net shaped covers, after capsule removal by pickling and heat treatment, before final machining.

### 5.2 Fabrication of bimetallic structures by HIP assisted diffusion bonding and explosion bonding

HIP assisted diffusion bonding might be a valid alternative to the assembly by brazing of cooling pipes onto structures that require cooling. For this purpose, an experimental study of the joining of different cuprous materials to stainless steel has been carried out by CERN in cooperation with the Fraunhofer Institute for Manufacturing Technology and Advanced Materials (IFAM) in Germany. This study consisted in the fabrication of a cooling system demonstrator using the HIP process to embed 316L stainless steel tubes into cuprous blocks [92]. Straight 316L steel tubes were embedded into machined blocks of 100 mm × 100 mm × 20 mm of different coppers (Cu-OF half hard; CuCr1Zr, hard drawn, and the ODS copper CEP DISCUP C3/30) through a solid-to-solid HIP assisted diffusion bonding process. A campaign of non-destructive and destructive examinations has proven that the interface between the stainless steel pipes and the cuprous blocks shows an outstanding continuity, with no major imperfections found at the tube-Cu interface [93].

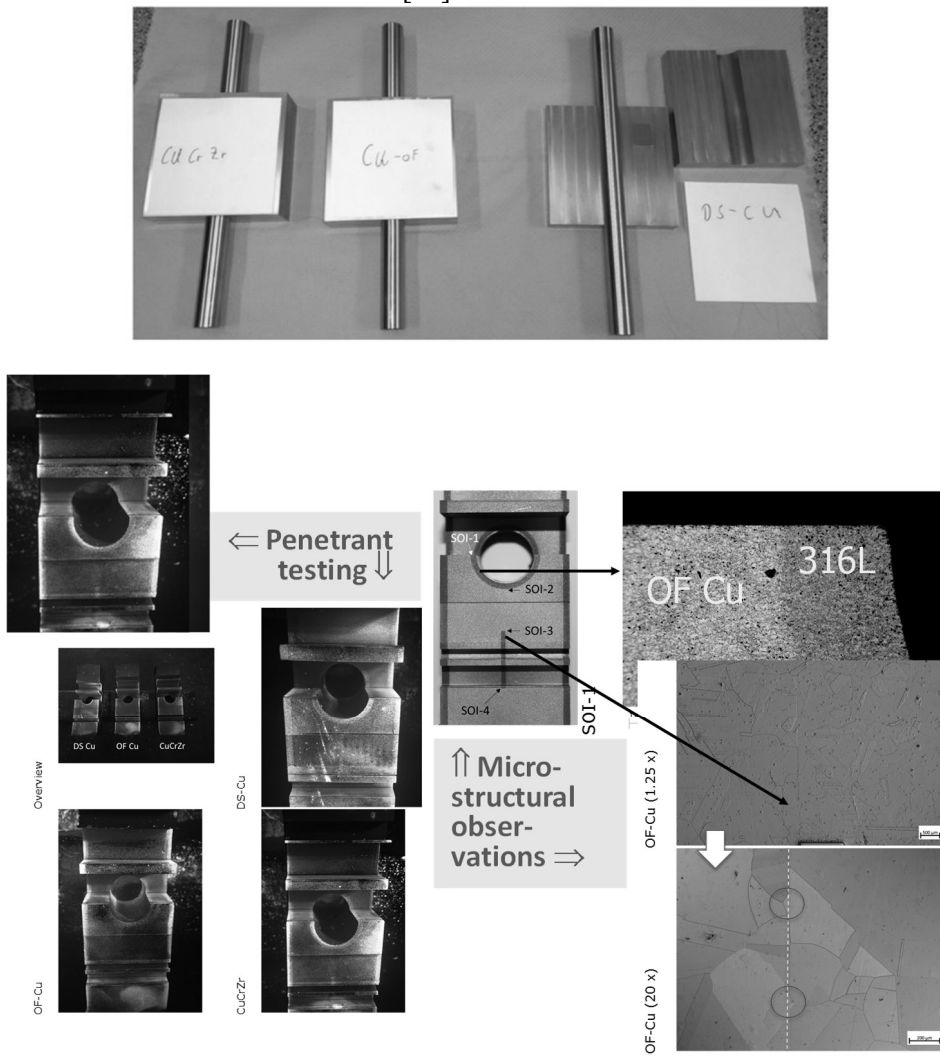

**Fig. 20**: Top: samples prepared for HIP diffusion bonding (from Ref. [92]). From left to right: CuCr1Zr, Cu-OF, and CEP DISCUP C3/30. The samples underwent a 950°C – 3 h – 100 MPa HIPing cycle. Bottom: examples of NDT by Penetrant Testing (PT) and destructive examinations by microoptical investigations of the Cu-stainless steel and Cu-Cu interface as issued from the HIPing diffusion bonding process. PT testing was performed according to the ISO 3452-1 with an Ardrox fluorescent 9703 penetrant and an Ardrox 9D1B developer. Images were taken after 30 minutes. Only a few indications were present [93].

Figure 20 shows the samples prepared for HIPing, together with examples of results of non-destructive and destructive examinations.

Explosion bonding between electrical coppers and stainless steel has also been successfully and extensively used for ITER electrical joints. These joints rely on the structural and the conductive properties of the bimetallic products, as well as on the tightness and strength of the interface between the conductive copper cladding and the structural stainless steel, which has to be tight to supercritical helium at 4.5 K [94].

## 5.3 Advanced NDT investigation techniques: X-ray Computed Micro-Tomography (CT) [95]

Figure 21 shows an example of CT applied to the helium inlets of the ITER correction coils (CCs). The CCs rely on a Cable-In-Conduit Conductor (CICC), whose supercritical cooling circuit at 4.5 K includes helium inlets and outlets. The assembly of the nozzles to the stainless steel conductor conduit involves manual fillet welds requiring full penetration through the thickness of the nozzle.

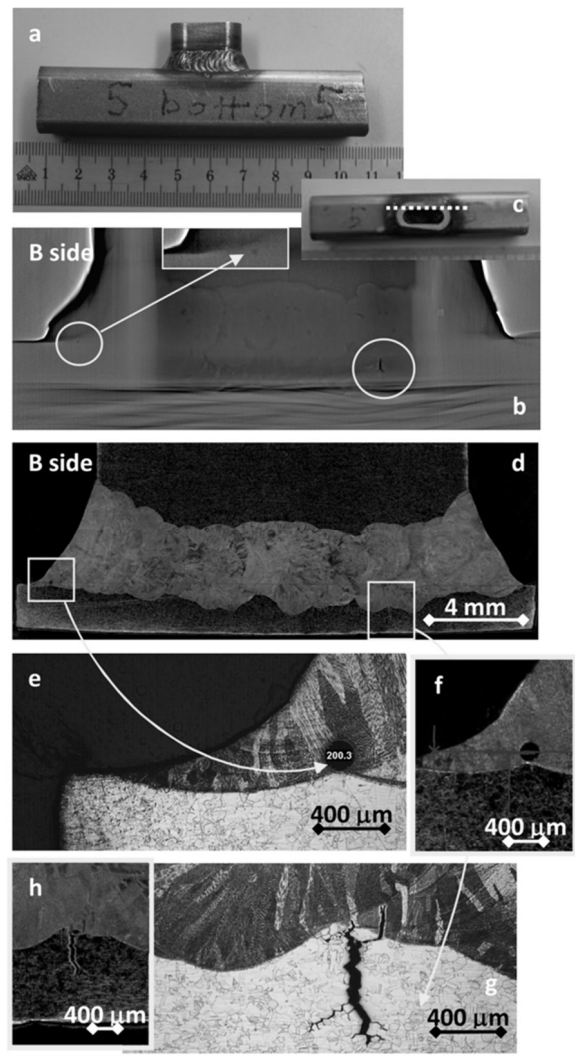

**Fig. 21**: Inlet (a) submitted to non-destructive followed by destructive inspections. A single view (b) featuring two imperfections in the same plane has been exported from the recorded tomographic inspection of the inlet. The view is relative to a laminographic 'cut' of the weld (dashed arrow in c). Porosity and crack are identified. The metallographic section (d) coincides with the laminographic cut in (b). Porosity (e and f) and cracks (g and h) are confirmed exactly in the plane and the positions identified by CT prior to cutting (from [95]).

Static and cyclic stresses have to be sustained by the inlet welds during operation. Severe constraints are imposed during welding, both in terms of position (welding on vertical conductor) and of the maximum allowed temperature not to be exceeded on the superconducting NbTi strands during welding. Moreover, those welds operating at 4.5 K must meet stringent quality levels of weld imperfections according to the standards in force. The entire volume of helium inlet and outlet welds is virtually uninspectable with sufficient resolution by conventional non-destructive examination techniques, in particular by conventional or computed radiographic testing or by UT. On the other hand, CT was successfully applied to inspect the full weld volume of several dozens of helium inlet qualification samples. The extensive use of CT techniques enabled significant progress in the weld quality of the CC inlets [95]. CT is a useful technique for inspection of welds, such as fillet welds, for which other techniques do not allow a 100 % volumetric inspection. Figure 21 also shows the impressive one-to-one correspondence between the tomographic cuts, which allow the thinnest imperfections such as planar defects to be detected in the volume, and micro-optical observations in the same positions. Other techniques such as Phased Array Ultrasonic Testing (PAUT) facilitate the volumetric inspection of complex shape components or welds [96].

# 6    Conclusions

The description of several applications of conventional and special metals and alloys for high vacuum purposes shows that a material not only consists of a 'chemical composition' or a designation, but is also the result of a complete metallurgy and metalworking process including possible refinement and remelting steps. A material for a demanding application is also defined through specified non-destructive and destructive tests, and should be ordered and delivered with a complete certification. In some cases, a quality control plan should be defined between the producer and the customer and should be followed during the production. The expected cost for a material adapted to an envisaged application will depend on an adequate specification defined before the order. Taking as an example austenitic stainless steels, a general purpose 304L, issued from a simple electric furnace primary melting, can have a cost that is 30 times lower than an ESR product multidirectionally forged to a particular final size and properly specified for a vacuum application.

The severity of the specification will depend on the final application. Whenever leak tightness has to be guaranteed across a thin wall (see the example of the VELO window of the LHC and experimental area interconnection components), an adequate microstructure in terms of grain size and inclusion content has to be specified. A fabrication resulting in a proper texture (orientation of the fibres parallel to the walls, or if this is not achievable, absence of orientation through multidirectional forging) should be preferred.

In modern accelerator construction, several components have to be integrated in a vacuum system through delicate welding operations, introducing internal stresses that can bring to leak a previously tested leak tight component. Advanced inspection techniques such as PAUT or CT are available, extending the possibilities of conventional NDT. In particular, CT might allow for a 100% volumetric inspection of welds otherwise virtually uninspectable by conventional NDT.

Applications at cryogenic temperatures, possibly associated with a requirement of non-magnetism, demand special care. A stainless steel, such as 316LN, specified as fully austenitic in order to be non-magnetic at RT will show, at cryogenic temperature, a higher magnetic susceptibility that might be unacceptable for some components. For the beam screen of LHC, a special austenitic stainless steel had to be developed to maintain low magnetic susceptibility in the base metal and the welds, in the temperature range between 10 K and 20 K. Strength and ductility requirements should not only be considered for the working temperature of a component, but defined as a function of all of the temperature cycles undergone by the vacuum component (NEG activation, baking, etc.).

Proper selection of materials should take into account their availability (e.g., EN AW-2219 is rarely available, special grades of stainless steels might require more than one year of lead time for delivery).

Corrosion issues should not be neglected, especially for thin walled components (see the example of the LHCb bellows). The use of halogen-activated fluxes should be strictly avoided in a stainless steel environment [97].

Material aspects should always be considered at the beginning of a project: issues of availability, interest in considering in a timely manner alternative techniques, including near-net shaping, as often late selections make the project more costly or incompatible with the tight schedule of an accelerator project. New projects will eventually result in less conservative solutions (bimetals by HIP assisted diffusion bonding or explosion bonding).

## Acknowledgements

The author wishes to thank the organizers of the special school on Vacuum for Particle Accelerators for the invitation to contribute and L. Akhouay for providing updated material prices.